\documentclass{aa}
\usepackage[varg]{txfonts}
\usepackage{graphicx}

\begin{document}

\title{Variability of the symbiotic X-ray binary GX 1+4:}

\subtitle{Enhanced activity near periastron passage}

\author{Krystian I\l{}kiewicz$^1$ \and Joanna Miko\l{}ajewska$^1$ \and Berto Monard$^2$} 

\offprints{K. I\l{}kiewicz, \email{ilkiewicz@camk.edu.pl}}

\institute{Nicolaus Copernicus Astronomical Center, Polish Academy of Sciences, Bartycka 18, 00716 Warsaw, Poland \and Kleinkaroo Observatory, Calitzdorp, Western Cape, South Africa}

\date{Received ... / Accepted ...}

\abstract {GX 1+4 belongs to a rare class of X-ray binaries with red giant donors,  symbiotic X-ray binaries. The system has a history of complicated variability on multiple timescales in the optical light and X-rays. The nature of this variability remains poorly understood.} { We study variability of GX 1+4 on long time-scale in X-ray and optical bands.} {The presented X-ray observations are from INTEGRAL Soft Gamma-Ray Imager and  RXTE All Sky Monitor. The optical observations are from INTEGRAL Optical Monitoring Camera.} {The variability of GX 1+4 both in optical light and hard X-ray emission (>17 keV) is dominated by $\sim$50--70d quasi-periodic changes. The amplitude of this variability is highest during the periastron passage, while during the potential neutron star eclipse the system is always at minimum, which confirms the 1161d orbital period that has had been proposed for the system based on radial velocity curve. Neither the quasi-periodic variability or the orbital period are detected in soft X-ray emission (1.3--12.2~keV), where the binary shows no apparent periodicity.} {}

\maketitle

\section{Introduction}

\object{GX 1+4} = \object{V2116 Oph} is an X-ray binary that was one of the brightest X-ray objects known at time of discovery \citep{1971ApJ...169L..17L}. It is harboring a pulsar with a spin period of $\sim$2~min that experienced one of the fastest spin-ups ever recorded \citep[e.g.][]{1981ApJ...243..257D,1989PASJ...41....1N}. The star reversed to a spin-down during an X-ray minimum in the early 1980's. This general spin-down trend with some spin-up episodes continues to date despite the restored high X-ray flux \citep[see][and references therein]{2012A&A...537A..66G}.

Using infrared observations \citet{1973NPhS..245...39G}  identified the optical counterpart of \object{GX 1+4}. The object showed spectral features typical for a symbiotic star (SySt), i.e. a binary system with a red giant (RG) donor \citep{1977ApJ...211..866D}. The symbiotic nature of \object{GX 1+4} was confirmed by discovery of flickering from the optical counterpart \citep{1997ApJ...482L.171J}. This makes the system a member of a rare class of symbiotic X-ray binaries (SyXB;  \citealt{2006A&A...453..295M}).

Lower limit of \object{GX 1+4} orbital period was discussed by \citet{1997ApJ...489..254C} to be $\ga$100d and most likely even $\ga$260d. In the case of elliptical orbit the period of binary can be estimated thanks to enhancement of mass transfer at periastron. Using this technique  \citet{1986ApJ...300..551C} and \citet{1999ApJ...526L.105P} found evidence of a $\sim$304d period in X-ray observations.  On the other hand no evidence of orbital period was found in Swift  Burst Alert Telescope (BAT) and the Rossi X-ray Timing Explorer (RXTE) All-Sky Monitor (ASM) observations \citep{2008ApJ...675.1424C}. \citet{1995AdSpR..16..131S} measured variations of H$\alpha$ emission line and concluded that the orbital period has to be significantly longer that the $\sim$300d period found in X-ray and optical observations. Recently, \citet{Majczyna_PTA} have analyzed optical $I$ light curve made by the OGLE IV project, and found a quasi-periodic variations with a timescale of 50--75 days, which they have attributed to changes in the accretion rate. They also found a strong peak in the power spectrum corresponding to the period of 295$\pm$70d, similar to the periods reported by \citet{1986ApJ...300..551C} and \citet{1999ApJ...526L.105P}, although they also noted that this result can be affected by substantial seasonal gaps in their light curve.

Finally, \citet{2006ApJ...641..479H}, based on high-resolution near-infrared spectroscopy spanning nearly 5 yr, determined the first spectroscopic orbit of the red giant.  They found an orbital  period of 1161$\pm$12d and modest eccentricity e=0.10$\pm$0.02 which makes \object{GX 1+4} the X-ray binary with longest orbital period by far known. The large mass function, $f(m)=0.371\pm0.026$ is consistent with the secondary being a neutron star (NS), and the red giant is the less massive component ($\la 1.3 \rm M\sun$).
\citet{2012A&A...537A..66G} showed that the spin-up of the NS is marginally correlated with periastron passage times calculated by \citet{2006ApJ...641..479H}.

Here we report analysis of \object{GX 1+4} variability on long time scales and, in particular, a search for orbitally related changes. We present our optical and X-ray observations in Sec.~\ref{obs_sec}. The results are in Sec.~\ref{res_sec} and conclusions in Sec.~\ref{conlc_sec}.

\begin{figure*}
  \resizebox{\hsize}{!}{\includegraphics{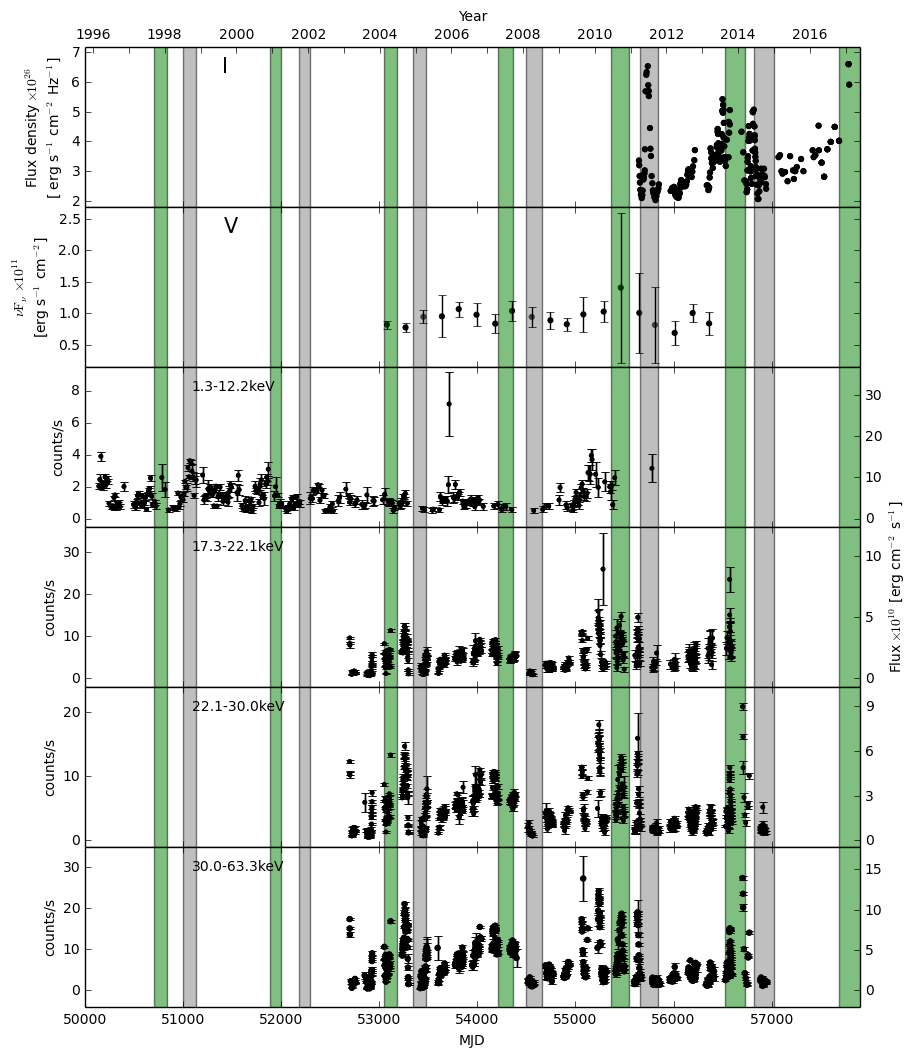}}
  \caption{Light curves of GX 1+4 in different spectral ranges. The green areas indicate time of periastron passage (Eq.~\ref{eq1}) and the gray areas indicate time of the potential eclipse of the neutron star (Eq.~\ref{eq2}). }
  \label{X-ray_var}
\end{figure*}

\section{Observations}\label{obs_sec}

We carried out optical observations of \object{GX 1+4} with a 35cm Meade RCX400 telescope in Kleinkaroo Observatory mounted with a SBIG ST8-XME CCD camera. The observations were carried out in a $Ic$ filter. Each observation was the result of several (3 to 6) individual exposures, which were calibrated (dark-subtraction and flatfielding) and stacked selectively. Magnitudes were derived from differential photometry to nearby reference stars using the single image mode of AIP4 image processing software. The photometric precision of the derived $Ic$ magnitudes is estimated to be better than 0.1 mag. Moreover, we supplied our photometry  with observations in the $I$-band scanned from \citet{Majczyna_PTA}. We shifted our observations so that the mean flux during the period covered both by ours and \citet{Majczyna_PTA} observations would be equal. This was done in order to ensure that the same photometric zeropoint was adapted in both datasets. Afterwards, we transformed the magnitudes to flux density using calibration of \citet{1979PASP...91..589B}. The observations are presented in Table~\ref{table:Iobs}, Figs.~\ref{X-ray_ogle} and \ref{X-ray_ogle}.

We gathered optical observations in $V$ filter from International Gamma Ray Astrophysics Laboratory (INTEGRAL) Optical Monitoring Camera (OMC; \citealt{2003A&A...411L.261M}), a refractive telescope with a 50~mm aperture. The telescope is equipped with a Johnson~$V$ filter, has 1024x1024 pixels imaging area CCD and 5$\,^{\circ}$x5$\,^{\circ}$ field of view. GX 1+4 is already listed in the INTEGRAL-OMC catalogue of optically variable sources \citep{2012A&A...548A..79A}. In our study we binned the observations with the bin size of 100d.

The collected soft X-ray observations are from the All Sky Monitor (ASM) on-board of the Rossi X-ray Timing Explorer (RXTE). The extracted lightcurve is from a spectral range 1.3--12.2~keV. Due to low signal to noise ratio in these observations we binned the data with 30d bins, and we excluded from our analysis data points  with signal to noise ratio $<3$. The ASM data was partially analyzed by  \citet{2008ApJ...675.1424C}, who did not find evidence of variability related to the orbital motion.

Hard X-ray data is from the INTEGRAL Soft Gamma-Ray Imager (ISGRI; \citealt{2003A&A...411L.141L}). This is a low energy detector of the Imager on Board the INTEGRAL Satellite (IBIS; \citealt{2003A&A...411L.131U}). The employed observations were from spectral ranges 17.3--22.1~keV, 22.1--30.0~keV and 30.0--63.3~keV. The observations are binned with a bin size of 1d. Data points with signal to noise ratio $<3$ were excluded from the analysis. The INTEGRAL ISGRI data was partly analyzed by \citet{2007A&A...462..995F} and \citet{2012A&A...537A..66G}.

The X-ray observations were transformed from count rates to physical flux. This was carried out first by transforming them to Crab units. The unit of one Crab was measured by taking a mean count rate of Crab during the period covered by each of the X-ray instruments. The observations were there transformed to cgs units using integrated flux of Crab in the corresponding X-ray bands. The flux of Crab was calculated using XSPEC v12.9.0 \citep{1996ASPC..101...17A} model of Crab spectrum with model parameters from \citet{2016MNRAS.456..775Z}.

All of the INTEGRAL and RXTE observations were extracted from the High-Energy Astrophysics Virtually ENlightened Sky database\footnote{www.isdc.unige.ch/heavens} \citep[HEAVENS;][]{2010int..workE.162W}. The light curves are presented in Figs~\ref{X-ray_var}~and~\ref{X-ray_ogle}.

\section{Results}\label{res_sec}

\begin{figure*}
  \resizebox{\hsize}{!}{\includegraphics{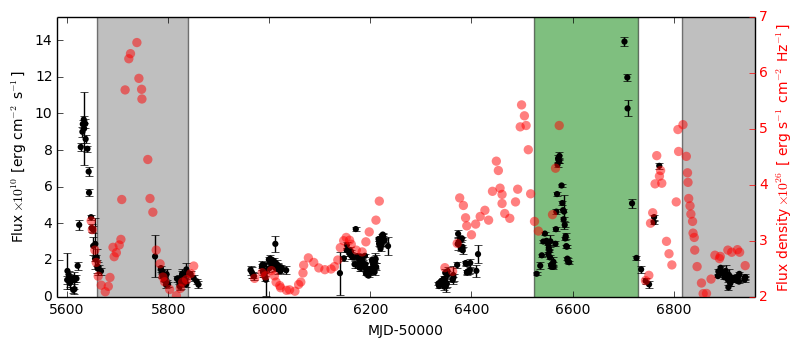}}
  \caption{Example of variability on short time-scale in 30.0--63.3~keV range (black points; left axis) together with observations in the $I$-band (red points; right axis) scanned from \citet{Majczyna_PTA}. The green area indicates time of periastron passage (Eq.~\ref{eq1}) and the gray areas indicate time of the potential eclipse of the neutron star (Eq.~\ref{eq2}).}
  \label{X-ray_ogle}
\end{figure*}

\begin{figure}
  \resizebox{\hsize}{!}{\includegraphics{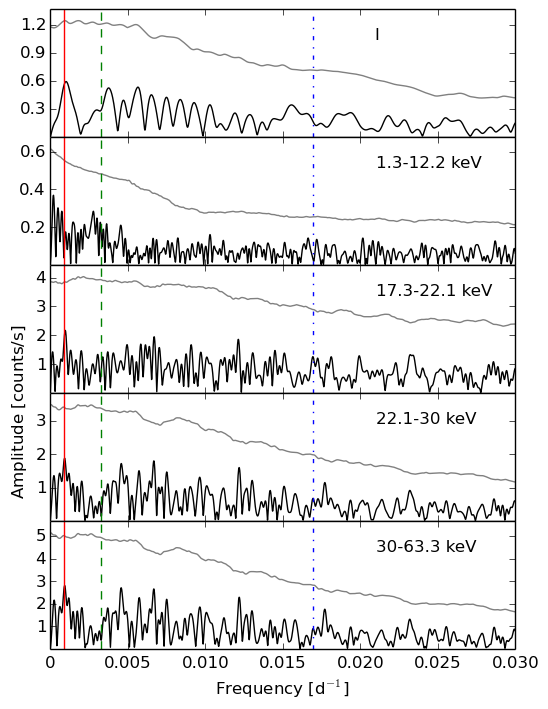}}
  \caption{Power spectrum of X-ray observations of \object{GX 1+4} (black lines). Solid red line represents orbital period of 1160.8d \citep{2006ApJ...641..479H}. Dashed green line represent proposed period of 303.8d \citep{1999ApJ...526L.105P}. Blue dashed-dotted line represents 59d period reported by \citet{Majczyna_PTA}. The gray line represents limit of a significant detection (see text).}
  \label{GX+4_fourier}
\end{figure}

Fourier analysis of the observations was performed using the discrete Fourier transform method in the $Period04$ program \citep{2005CoAst.146...53L}. The resulting power spectra are presented in Fig.~\ref{GX+4_fourier}. The power spectra are apparently dominated by a low-frequency noise. In order to estimate the significance level of peaks in the power spectrum we assumed the noise level to be equal to an average amplitude in the interval $\pm$0.005$\mathrm{d}^{-1}$ to a given frequency. We constructed significance curve adopting signal to noise ratio $>4$ as a limit of significant detection, as was suggested e.g. by \citet{1993A&A...271..482B}. 

\begin{figure}
  \resizebox{\hsize}{!}{\includegraphics{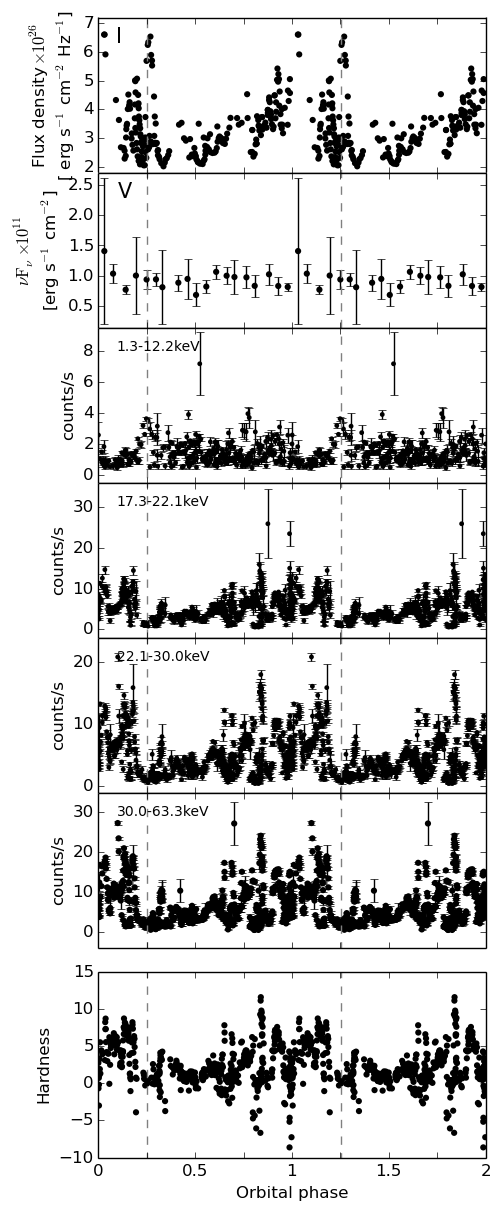}}
  \caption{Top five panels: phase plot of GX 1+4 at different spectral ranges with the orbital period of 1160.8d \citep{2006ApJ...641..479H}, where orbital phase $\phi=0$ corresponds to the time of periastron passage (Eq.~\ref{eq1}). The vertical grey lines are times of neutron star eclipse according to Eq. (\ref{eq2}). Bottom panel: phase plot of hardness  of GX 1+4. The hardness is defined as a difference between count rate in 30.0--63.3~keV range and in 17.3--22.1~keV range.}
  \label{X-ray_phase}
\end{figure}

\begin{figure}
  \resizebox{\hsize}{!}{\includegraphics{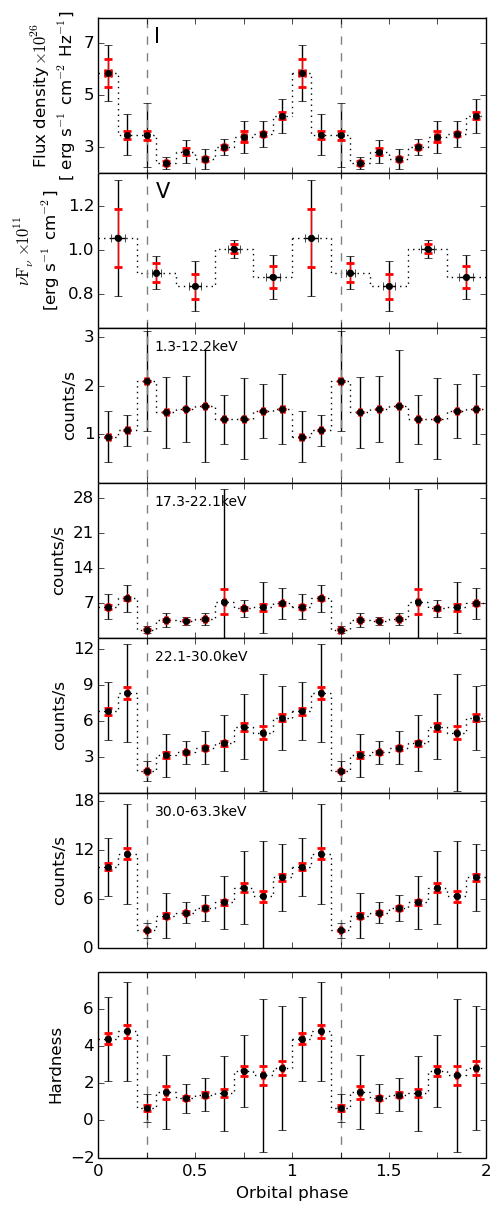}}
  \caption{Same as Fig.~\ref{X-ray_phase}, but with binned data points. The top panels has been created with bin size of 0.2 of orbital phase. In case of other panels bin size of 0.1 of orbital phase was employed. The red error bars represent standard error of mean. The black error bars represent standard deviation.}
  \label{X-ray_phase_bin}
\end{figure}

Neither the 1160.8d orbital period found by \citet{2006ApJ...641..479H} nor the 
303.8d period reported by  \citet{1999ApJ...526L.105P}, and the 59d period 
present in the optical observations of \citet{Majczyna_PTA} were clearly 
detected in the the power spectra in Fig.~\ref{GX+4_fourier} with a statistical 
significance. However, the X-ray light curves in the 17.3-63.3~keV range (Fig.~\ref{X-ray_var}) consists of 
series of flares with characteristic timescales of 50--70d similar to the 
quasi-periodic changes reported by \citet{Majczyna_PTA}, and there is no permanent X-ray emission. The comparison of the 
ISGRI light and $I$ light curves  (Fig.~\ref{X-ray_var},\ref{X-ray_ogle}) shows 
that virtually the same phenomenon is observed in the hard X-rays and the 
optical $I$ light. The quasi-periodic nature of flares caused a low frequency noise in the power spectra, because of which the orbital period was not formally detected. However, the flares seem to be particularly prominent when the binary approaches periastron. \citet{Majczyna_PTA} suggested that optical brightening with low and high amplitude could be caused by a different mechanism.  Using the hard X-ray observations (>22.1~keV) with longer baseline it is clear that the high amplitude flares occur only close to the periastron passage, i.e. when enhanced mass transfer rate can be expected. 
The correlation between the flare activity and the orbital position is better visible in the phased light curves (Fig.~\ref{X-ray_phase} and Fig~\ref{X-ray_phase_bin}) which  were generated using ephemeris
\begin{equation}\label{eq1}
T=~JD~2,451,943(\pm53)\pm1160.8(\pm12.4)\times E
\end{equation}
for the periastron passage and ephemeris 
\begin{equation}\label{eq2}
T=~JD~2,452,236(\pm53)\pm1160.8(\pm12.4)\times E
\end{equation}
for the potential eclipse of the neutron star \citep{2006ApJ...641..479H}. 

Moreover no brightenings in X-rays are observed around the time of possible NS eclipse, which leads to skewed phase plot of hard X-ray observations (Fig.~\ref{X-ray_phase_bin}). The variations in amplitude of quasi-periodic changes confirms the orbital period derived by  \citet{2006ApJ...641..479H} and indicates that the big and low amplitude brightenings are the same phenomenon with different amplitude at different orbital phase. We note that one of the flares in $I$ band was observed near the time of possible NS eclipse (Fig.~\ref{X-ray_ogle}), however the uncertainty of the time of potential eclipse is larger than the timescale of the flare, because of accumulation of errors in ephemeris of \citet{2006ApJ...641..479H} over few cycles. 
In fact, increasing the orbital period by $\sim 1\,\sigma$ and/or somewhat later periastron passage would remove the problem with appearance of flares during the eclipse.  Alternatively if flares during the NS eclipse would be observed in the optical light only and not in X-rays this could indicate that the  source of the optical light is at large distance from the NS  (e.g. in jets), and it is not eclipsed.

Enhanced mass transfer rate at periastron as a source of bigger amplitude of the flares is consistent with the fact that the maximum flux is observed at orbital phase  $\phi\sim0.2$ (where $\phi=0$ is the time of the periastron passage; Fig.~\ref{X-ray_phase_bin}), which is expected given that the binary is relatively wide and the matter from RG wind needs time to reach the NS. The orbital variability is more prominent in the harder range, which can be observed as hardening of the spectrum close to the periastron passage (Fig.~\ref{X-ray_phase_bin}). This indicates that the harder spectral range may be better for searching for orbital period in other SyXBs.

Maxima of short-term variations in  17.3-63.3~keV range seem to occur at the same times as maxima of variability in the optical ($I$) range (see Fig.~\ref{X-ray_ogle}; \citealt{Majczyna_PTA}). Similar correlation in GX~1+4 was found in the past between X-ray flares, observed by the BATSE experiment on the Compton Gamma-Ray Observatory (CGRO), and variability of the  H$\alpha$ emission line \citep{1995AdSpR..16..131S,1995MNRAS.274L..59G}. It is worth noting that decline of both H$\alpha$ and hard X-ray emission from the July--September 1993 flare coincided to within a few days, however the flare in H$\alpha$ started $\sim$30d before the flare in X-rays. \citet{1995MNRAS.274L..59G} noticed that this points to the fact that both X-rays and H$\alpha$ variability has the same source, however H$\alpha$ variability could not be directly caused by changes in X-ray radiation. If the brightening is caused by a short increase in the mass accretion rate trough the accretion disc, the time lag can be estimated using mass of the NS, M$_{\mathrm{NS}}$, mass accretion rate, $\dot{M}$, viscosity, $\alpha$, and the effective temperature corresponding to the maximum in the H$\alpha$ emission, $T_{\mathrm{eff,H\alpha}}$ \citep{2016arXiv160507013B}. After adopting the total X-ray luminosity L$_{\mathrm{X}}\simeq4\times10^{37}$~erg~s$^{-1}$ \citep[e.g.][]{2001MNRAS.325..419G} and assuming the typical parameters of the NS, $M_{\mathrm{NS}}$=1.35~M$_\odot$ \citep{1999ApJ...512..288T} and radius $R_{\mathrm{NS}}$=10~km, the mass accretion rate can be estimated from the equation $\dot{M}=L_\mathrm{X}\,R_{\mathrm{NS}}/(G\,M_{\mathrm{NS}})$ to be $\sim$3.5$\times10^{-9}$~M$_\odot$~yr$^{-1}$. Assuming $T_{\mathrm{eff,H\alpha}}$=10000K, corresponding to a typical ionization temperature of hydrogen in a symbiotic nebula, and $\alpha$=0.2 we estimate the time delay between the observations in X-ray range in H$\alpha$ emission to be $\sim$28d, in good agreement with the observations. 

We emphasize that there is not any detectable time lag between the optical $I$ and X-ray flares, hence, in contrary to the H$\alpha$ emission,  the optical $I$ probably has a different origin than the increased mass transfer rate trough the outer parts of accretion disc. Although the timescales of the semi-periodic $I$-band variability is similar to those of semi-regular (SR) pulsations of red giants, such SR pulsations have much lower amplitudes, $\la 0.2$ mag as compared to $\Delta I$ up to 1 mag (Fig.~\ref{X-ray_ogle}; \citealt{Majczyna_PTA}), and their periods are well defined
(e.g. \citealt{2013AcA....63..405G}). Moreover, this variability practically vanishes during inferior conjuctions when the red giant should dominate the $I$-band light. 
On the other hand, similar flares on time scale of tens of days have been observed simultaneously in X-rays, optical and near infrared emission in some accreting low-mass X-ray binaries \citep[LMXB; see e.g][and references therein]{2013MNRAS.430.3196V}. While they seem to be due to unstable disk-accretion, the nature of the optical and near infrared emission remains poorly understood. In particular, it has been suggested that these bands are dominated by either the jet emission,  extended hot
accretion flow or outer irradiated part of the accretion disk heated by the X-rays.
Although most of the LMXB discussed by \citet{2013MNRAS.430.3196V} host black holes, their accretion disk behavior should not be much different from those with neutron stars. It is thus tempting to assume similar origin of the flared emission in GX 1+4; all these scenarios would produce optical radiation with much shorter delay time compared to the X-rays, which would not be detected in our observations. It is also worth to note that the X-ray and optical variability has similar relative amplitude (Fig.~\ref{X-ray_ogle}), which in future could be used to determine the origin of the optical radiation \citep{2006MNRAS.371.1334R}.

The sawtooth-wave like shape of the binned phase plot of observations in 17.3-63.3~keV range (Fig.~\ref{X-ray_phase_bin})  is somewhat unusual for an X-ray binary, given that it shows slow rise and fast decay, while the opposite is usually observed, i.e. fast rise, and slow decay. Another object with sawtooth-wave like variability skewed in the same way was  X-ray nova \object{GRS 1716-249} \citep{1996ApJ...470L.105H}. While variability of this object was not related to orbital motion, the sawtooth-wave was almost identical to the binned phase plot of observations in 17.3-63.3~keV range. Moreover, it was observed in similar spectral range, namely 20--100~keV. \citet{1996ApJ...470L.105H} suggested, based on their radio observations, that variability of \object{GRS 1716-249} was related to ejection of relativistic particles in a jet. \citet{1998ApJ...505..854E} proposed, that \object{GRS 1716-249} during the flare maxima was in the the low spectral state and the rapid decay was related to rapid increase in the mass accretion rate that precipitated a corresponding decrease in transition radius. During minima in the 20--100~keV range, \object{GRS 1716-249} would be in high energy state, and the transition radius would be close to the last stable orbit. Similar mechanism might take place in \object{GX 1+4}, given that the fast decline on binned phase-plot  fallows shortly after the periastron passage (Fig.~\ref{X-ray_phase_bin}), when increased mass transfer rate is expected. In that case, the rapid decline would only coincidentally be observed close to the time of spectroscopic conjunction. 

The binned phase-plot is a measure of amplitude of the $\sim$50--70d quasi-periodic variability. The averaged sawtooth-wave like variability could be related to the fact that the spectrum of quasi-periodic flares shifted towards harder spectral range than the observed one. This seems unlikely given that the observed hardness seems to be the lowest near the observed minimum (Fig~\ref{X-ray_phase_bin}). The more likely explanation is that the quasi-periodic flares originate in a region in accretion disc that would be affected by change in the transition radius. Moreover, it seems that the sawtooth-wave shape is not always present. In the lightcurve in 17.3-63.3~keV range the drop to minimum in years 2006-2007 is slower, resembling more a skewed sinusoidal function (Fig.~\ref{X-ray_var}). In the other observed drops in amplitude of the $\sim$50--70d quasi-periodic variability, in 2004-2006, 2010-2011 and 2013-2014, the variability indeed resembles a sawtooth-wave  (Fig.~\ref{X-ray_var}).

A marginally detectable variability seems to be present in the optical observations of GX 1+4 in $V$ filter   (Fig.~\ref{X-ray_var},~\ref{X-ray_phase_bin}). The variability may be combination of orbital period (Fig.~\ref{X-ray_phase_bin}) and short term variability discussed by \citet{Majczyna_PTA}, but quality of the data is not sufficient for any meaningful analysis. However the orbital period of 1160.8d is not excluded on basis of observations in $V$ filter, that have longer baseline than the observations in $I$ filter, which in turn show variability related to orbital motion (Fig.~\ref{X-ray_phase_bin}). 

The light-curve in 1.3--12.2~keV band (Fig.~\ref{X-ray_var}) shows irregular variability. This is consistent with the fact that \citet{2008ApJ...675.1424C} did not detect orbital period on basis of the same observations, although with shorter baseline. The lack of detectable orbital variability may be due to the fact that the variability in this range is dominated by variable column density rather than variable mass transfer rate, as was observed e.g. in \object{RT Cru} and suggested for \object{T~CrB} (\citealt{2009ApJ...701.1992K,ilkiew16}). It is interesting to note that near the orbital phase $\phi\sim0.1$ there is a minimum in the phase plot of the observations in the 1.3--12.2~keV band (Fig.~\ref{X-ray_phase}). This minimum is clearly visible in the binned phase plot (Fig.~\ref{X-ray_phase_bin}) and it occurs at the same orbital phase maximum in harder X-rays. This may be a coincidence since variability in 1.3--12.2~keV and  17.3--63.3~keV bands is not clearly correlated (Fig.~\ref{X-ray_var}) and the orbital period is not detected in the power spectrum of the observations in 1.3--12.2~keV range (Fig.~\ref{GX+4_fourier}). However is is possible that this is due to pivoting of  the X-ray  spectral energy distribution.

\section{Conclusions}\label{conlc_sec}

In this work we analyzed long term X-ray and optical variability of a SyXB \object{GX 1+4}. The main conclusions are:

\begin{itemize}
\item  There is not permanent hard X-ray (>17~keV) emission in \object{GX 1+4}. Both hard X-ray and optical $I$ band are dominated by flares occurring quasi-periodically on $\sim$50--70d timescale and showing no time-lag between the two spectral ranges. On the contrary, in the past, delay between X-ray and H$\alpha$ emission has had been observed.
\item The observations in the hard X-rays and optical $I$ band confirm the orbital period of 1160.8d derived on basis of IR spectroscopy by \citet{2006ApJ...641..479H}. In particular, we see pronounced, regular behavior in hard X-rays and optical $I$ band, while it is not visible in soft X-rays (Fig.~\ref{X-ray_phase_bin}).
\item The period of $\sim$300d measured on basis of periodic spin-up of the NS \citep{1986ApJ...300..551C,1999ApJ...526L.105P}  and optical photometry \citep{Majczyna_PTA} is not confirmed by observations in both soft (1.3-12.2~keV) and hard X-rays.
\item The variability in the soft X-ray band shows no apparent periodicity, but seems to be always at minimum during the maximum of hard X-rays. 
\item Detection of orbital period in hard X-rays and no-detection in soft X-rays shows that the former may be better to search for orbital period in other SyXB. This would explain why orbital period was not detected in \object{4U 1700+24} and \object{4U 1954+31} on basis of observations in softer X-rays \citep{2008ApJ...675.1424C}.
\end{itemize}

\object{GX 1+4} shows variability on multiple timescales and spectral properties of the object on any timescale are not uniquely determined (see \citealt{2007A&A...462..995F} and references therein). Therefore future X-ray and optical spectroscopy  monitoring of the system would be the next step towards understanding the variability of this SyXB. We note that the next spectroscopic conjunction will occur in October 2017.

\begin{acknowledgements} 
We are grateful to Andrzej Zdziarski for a helpful discussion. KI has been financed by the Polish Ministry of Science and Higher Education Diamond Grant Programme via grant 0136/DIA/2014/43. This study has been partially financed by Polish National Science Centre grants 2012/04/M/ST9/00780, and 2015/18/A/ST9/00746.

\end{acknowledgements}

\bibpunct{(}{)}{;}{a}{}{,} % to follow the A&A style
\bibliographystyle{aa} % style aa.bst
\bibliography{references} % your references Yourfile.bib

\begin{appendix}
\section{The 304d period}
\begin{figure}
  \resizebox{\hsize}{!}{\includegraphics{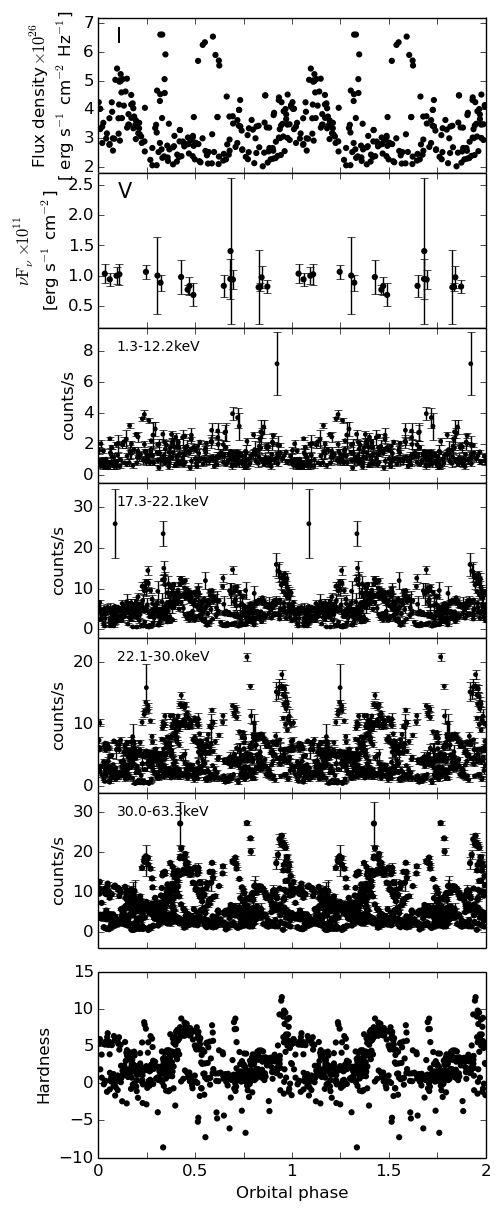}}
  \caption{Same as Fig.~\ref{X-ray_phase}, but with P=303.8d \citep{1999ApJ...526L.105P}.}
  \label{X-ray_phase_short}
\end{figure}

\begin{figure}
  \resizebox{\hsize}{!}{\includegraphics{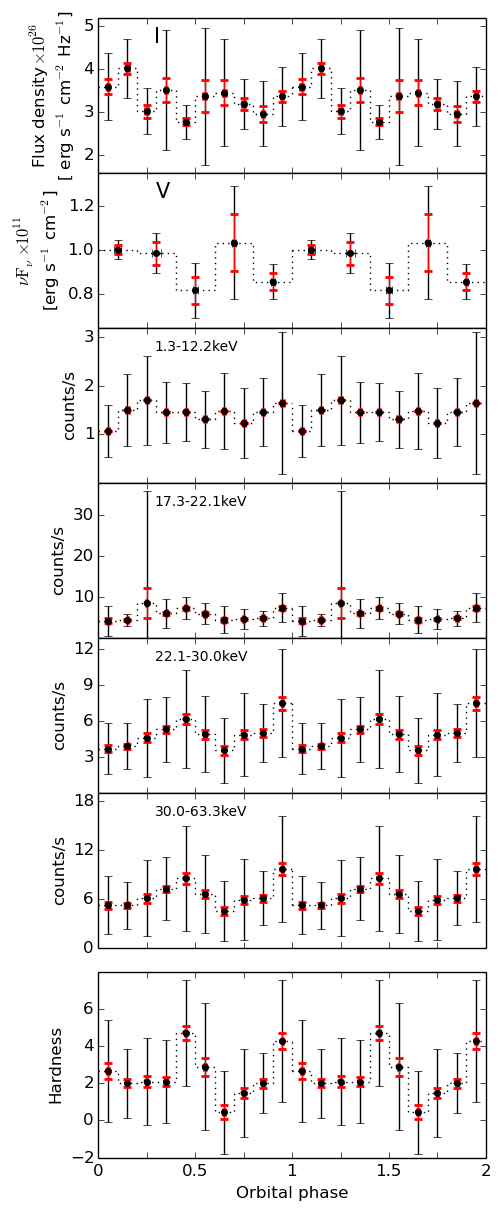}}
  \caption{Same as Fig.~\ref{X-ray_phase_bin}, but with P=303.8d \citep{1999ApJ...526L.105P}.}
  \label{X-ray_phase_short_bin}
\end{figure}

In this section we perform similar analysis as in the main body of the text, but for $\sim$304d period proposed on basis of X-ray observations \citep{1986ApJ...300..551C,1999ApJ...526L.105P}. In none of the observed wavelength bands we clearly detect correlation of our observations with this period (Fig.~\ref{X-ray_phase_short}). In particular, the amplitude of $\sim$50--70d quasi-periodic changes seems to be randomly distributed with respect to this period, with no clearly separated times of low and high amplitude flares, as in case of the orbital period (Fig.~\ref{X-ray_phase}). The binned phase plot of hard X-ray data (>22.1~kev) shows possible, small correlation with the $\sim$304d period (Fig.~\ref{X-ray_phase_short_bin}). If the amplitude of flares is related to the $\sim$304d period, the correlation is much weaker than in the case of the orbital period. This could be addressed in future studies, when more data will be available.

\section{Observational data}
\begin{table}
\caption{Observations of \object{GX 1+4} from the Kleinkaroo Observatory.}         
\label{table:Iobs}      
\centering                          
\begin{tabular}{c c c}        
\hline\hline                
MJD & $Ic$ [mag] & Flux density $\times10^{26}$ \\ 
    &            & [ erg s$^{-1}$ cm$^{-2}$ Hz$^{-1}$] \\    
\hline                       
56339.00 & 15.02 & 2.51\\
56355.00 & 15.09 & 2.35\\
56368.00 & 14.91 & 2.78\\
56386.00 & 14.68 & 3.43\\
56402.00 & 14.71 & 3.34\\
56414.00 & 14.69 & 3.40\\
56430.00 & 14.70 & 3.37\\
56440.80 & 14.57 & 3.80\\
56448.80 & 14.43 & 4.32\\
56472.70 & 14.72 & 3.31\\
56490.80 & 14.45 & 4.25\\
56502.80 & 14.28 & 4.96\\
56509.80 & 14.48 & 4.13\\
56517.80 & 14.67 & 3.47\\
56527.80 & 14.67 & 3.47\\
56542.80 & 14.50 & 4.05\\
56561.80 & 14.35 & 4.65\\
56573.80 & 14.37 & 4.57\\
56692.10 & 14.43 & 4.32\\
56710.10 & 14.62 & 3.63\\
56722.00 & 14.95 & 2.68\\
56738.00 & 14.96 & 2.65\\
56743.90 & 15.00 & 2.56\\
56752.90 & 14.82 & 3.02\\
56756.10 & 14.68 & 3.43\\
56772.00 & 14.49 & 4.09\\
56786.90 & 14.77 & 3.16\\
56806.00 & 14.47 & 4.17\\
56806.90 & 14.27 & 5.01\\
56825.05 & 14.47 & 4.17\\
56845.90 & 14.87 & 2.88\\
56867.90 & 15.02 & 2.51\\
56884.90 & 14.98 & 2.61\\
56892.85 & 14.80 & 3.08\\
56900.85 & 14.87 & 2.88\\
56921.80 & 14.80 & 3.08\\
56943.80 & 15.07 & 2.40\\
57070.10 & 14.67 & 3.47\\
57085.10 & 14.65 & 3.53\\
57100.05 & 14.83 & 2.99\\
57111.15 & 14.86 & 2.91\\
57136.10 & 14.84 & 2.96\\
57162.00 & 14.96 & 2.65\\
57187.95 & 14.66 & 3.50\\
57217.80 & 14.82 & 3.02\\
57226.75 & 14.93 & 2.73\\
57240.75 & 14.83 & 2.99\\
57255.80 & 14.78 & 3.13\\
57299.70 & 14.69 & 3.40\\
57323.75 & 14.83 & 2.99\\
57418.10 & 14.60 & 3.70\\
57428.10 & 14.67 & 3.47\\
57445.10 & 14.65 & 3.53\\
57478.15 & 14.38 & 4.53\\
57481.05 & 14.60 & 3.70\\
57508.95 & 14.73 & 3.28\\
57534.84 & 14.90 & 2.80\\
57567.93 & 14.59 & 3.73\\
57601.82 & 14.52 & 3.98\\
57642.82 & 14.39 & 4.49\\
57686.78 & 14.51 & 4.02\\
57782.12 & 13.97 & 6.61\\
57785.10 & 13.97 & 6.61\\
57790.12 & 14.09 & 5.91\\

\hline                                 
\end{tabular}
\end{table}
\end{appendix}
\end{document}